\documentclass[conference]{IEEEtran}
\IEEEoverridecommandlockouts

\usepackage[english]{babel}
\usepackage{blindtext}
\usepackage{tikz}
\usepackage{amsmath}
\usepackage{tabu}
\usepackage{adjustbox}
\usepackage{filecontents}
\usepackage{xspace}
\usepackage{multirow}
\usepackage{multicol}
\usepackage{subcaption}
\usepackage{titlesec}
\usepackage[font=small]{caption} 

\titlespacing*{\section}
{0pt}{1.5ex}{2.0ex}

\renewcommand{\baselinestretch}{0.958}   
\usepackage{tikz}
\usepackage{amsmath}

\usepackage{multicol}
\usepackage{comment}
\usepackage{float}
\usepackage{csquotes}
\usepackage{textcomp}

\usepackage{filecontents}
\usepackage{xspace}
\usepackage{multirow}

\usepackage{multicol}
\usepackage{lipsum}
\usepackage{mwe}
\usepackage{subcaption}

\definecolor{darkerblue}{RGB}{0,63,202}
\definecolor{sabzseyedi}{RGB}{14,109,25}
\definecolor{LightGray}{RGB}{247,247,247}

\newcommand{\algo}{TenFor\xspace}  
\newcommand{\storyline}{StoryLine View\xspace} 
\newcommand{\tableview}{Table View\xspace} 

\newcommand{\reminder}[1]{{\textsf{\textcolor{red}  {[#1]}}}}

\usepackage{pifont} 
\newcommand{\cmark}{\ding{51}}%
\newcommand{\xmark}{\ding{55}}%

\usepackage{array} 
\newcolumntype{P}[1]{>{\centering\arraybackslash}p{#1}}

\usepackage{fancyhdr}
\usepackage{kantlipsum}
\fancypagestyle{plain}{
\fancyhf{}
\fancyhead[C]{\textbf{2020 IEEE/ACM International Conference on Advances in Social
Networks Analysis and Mining (ASONAM)}} 

}
\usepackage{eso-pic}

\begin{document}

\AddToShipoutPictureBG*{
\AtPageUpperLeft{
\setlength\unitlength{1in}
\hspace*{\dimexpr0.5\paperwidth\relax}
\makebox(0,-0.75)[c]{\textbf{2020 IEEE/ACM International Conference on Advances in Social
Networks Analysis and Mining (ASONAM)}}}}

\title{\algo: A Tensor-Based Tool to Extract Interesting Events from Security Forums\\}

\author{
{\rm Risul Islam}\\
UC Riverside\\
risla002@ucr.edu
\and
{\rm Md Omar Faruk Rokon}\\
UC Riverside\\
mroko001@ucr.edu
\and
{\rm Evangelos E. Papalexakis}\\
UC Riverside\\
epapalex@cs.ucr.edu
 \and 
{\rm Michalis Faloutsos}\\ 
UC Riverside\\
michalis@cs.ucr.edu
} 

\maketitle

\IEEEoverridecommandlockouts
\IEEEpubid{\parbox{\columnwidth}{\vspace{8pt}
\makebox[\columnwidth][t]{IEEE/ACM ASONAM 2020, December 7-10, 2020}
\makebox[\columnwidth][t]{978-1-7281-1056-1/20/\$31.00~\copyright\space2020 IEEE} \hfill}
\hspace{\columnsep}\makebox[\columnwidth]{}}
\IEEEpubidadjcol

\newcommand{\risul}[1]{{\bf {\textcolor{green}{Risul:}}{\textcolor{blue}{#1}}}}

\newcommand{\rismi}[1]{{\bf {\textcolor{green}{R+M:}}{\textcolor{red}{#1}}}}

\newcommand{\miii}[1]{{\bf  {\textcolor{blue}{MF:}}{\textcolor{red}{#1}}}}

\definecolor{darkgreen}{RGB}{75,133,100}
 \newcommand{\miok}[1]{{\bf {\textcolor{darkgreen}{#1}}}}

\newcommand{\hide}[1]{}
\newcommand{\semihide}[1]{{\tiny #1}}
\newcommand{\changed}[1]{{\textcolor{blue}{#1}}}
\newcommand{\proofApp}[1]{{\emph{\bf Proof of #1.}}}
\newcommand{\mkclean}{
    \renewcommand{\reminder}[1]{}
    \renewcommand{\comment}[1]{}
    \renewcommand{\semihide}[1]{}
    \renewcommand{\changed}[1]{}
}

\newcommand{\RelThreads}{$R_t$}

\newcommand{\method}{{\tt HackerChatter}}
\newcommand{\source}{O}
\newcommand{\destination}{D}
\newcommand{\node}{\ensuremath{v}}
\newcommand{\edge}{\ensuremath{e}}
\newcommand{\graph}{\ensuremath{\mathcal{G}}}
\newcommand{\length}{\ensuremath{{l}}}
\newcommand{\prob}{\pi} 
\newcommand{\error}{e}
\newcommand{\fun}{f}
\newcommand{\loc}{\lambda}
\newcommand{\outdeg}{d_{out}}
\newcommand{\base}{b}
\newcommand{\dist}{\delta}
\newcommand{\ratio}{\rho}
\newcommand{\extrad}{\alpha}
\newcommand{\nodec}{\nu}
\newcommand{\pvalue}{p-value}
\newcommand{\mean}{average}
\newcommand{\std}{standard-deviation}
\newcommand{\ba}{\beta_{0}}
\newcommand{\bb}{\beta_{1}}
\newcommand{\bc}{\beta_{2}}

\newcommand{\HH}{Heavy-Hitters\xspace}
\newcommand{\CS}{Consumers\xspace}
\newcommand{\HP}{High Producers\xspace}
\newcommand{\AP}{Average Producers\xspace}
\newcommand{\LP}{Low Producers\xspace}
\newcommand{\PR}{\textit{Producers}\xspace}
\newcommand{\Rho}{\mathrm{P}}
\newcommand{\expnumber}[2]{{#1}\mathrm{e}{#2}}

\newtheorem{problem}{Problem}
\newtheorem{definition}{Definition}
\newtheorem{fact}{Fact}
\newtheorem{observation}{Observation}
\newtheorem{proposition}{Proposition}

\newcommand{\OffComm}{OffensiveCommunity\xspace}
\newcommand{\OffCommShort}{OffensComm.\xspace}
\newcommand{\HTS}{HackThisSite\xspace}
\newcommand{\WSForum}{WildersSecurity\xspace}
\newcommand{\Ash}{Ashiyane\xspace}
\newcommand{\WSShort}{WildersSec.\xspace}
\newcommand{\Ethical}{EthicalHackers\xspace}
\newcommand{\EthicalShort}{EthicHacks\xspace}
\newcommand{\darkode}{Darkode\xspace}

\newcommand{\TT}{{\em Hacks}\xspace}
\newcommand{\PS}{{\em Services}\xspace}
\newcommand{\AN}{{\em Alerts}\xspace}
\newcommand{\AG}{{\em Experiences}\xspace}

\newcommand{\MV}{{\em Malware - Virus}\xspace}
\newcommand{\HA}{{\em Hack - Account}\xspace}
\newcommand{\TG}{{\em Tutorial - Guide}\xspace}
\newcommand{\SB}{{\em Sell - Buy}\xspace}
\newcommand{\AV}{{\em Attack - Vulnerability}\xspace}
\newcommand{\VG}{{\em Video - Game}\xspace}
\newcommand{\CC}{{\em Credit Card}\xspace}

\newcommand{\myalg}{RKEA\xspace}
\newcommand{\myalgFull}{{\bf R}ecursive {\bf K}eywords {\bf E}xtracting  {\bf A}pproach\xspace}

\newcommand{\seedMethod}{Initialization via domain adaptation\xspace}

\newcommand{\seedMethodAcro}{IDA\xspace}

\newcommand{\WordsFreq}{\textit{TextInfo}\xspace}
\newcommand{\IPP}{\textit{DecimalVal}\xspace}
\newcommand{\Mixed}{\textit{Mixed}\xspace}
\newcommand{\Cocluster}{\textit{Co-Cluster}\xspace}

\newcommand{\extUserWord}{\textit{ContextInfo}\xspace}
\newcommand{\extUserWordlong}{Contextual Information\xspace}

\newcommand{\PostTextlong}{Text information of the post\xspace}   
\newcommand{\postTextlong}{text information of the post\xspace}
\newcommand{\postText}{\textit{PostText}\xspace}

\newcommand{\mfal}[1]{{\bf{\textcolor{blue}{MF:}}{\textcolor{red}{#1}}}}
\newcommand{\mrem}[1]{{\bf{\textcolor{green}{\xspace#1}}}}

\newcommand{\idenp}{Identification\xspace}
\newcommand{\classp}{Characterization\xspace}

\newcommand{\crosstrain}{cross-training\xspace} 

\newcommand{\cseed}{Cross-Seeding\xspace} 

\newcommand{\cporting}{Basic\xspace} 

\newcommand{\kwords}{$W$\xspace}
\newcommand{\kwordsDef}{Word-Range\xspace}
\newcommand{\norm}[1]{\left\lVert#1\right\rVert}

\newcommand{\eatreminder}{\renewcommand{\reminder}{\hide}}

\newcommand{\ThSim}{T_{sim}\xspace}  
\newcommand{\ThWord}{T_{key}\xspace}  
\newcommand{\mdim}{m\xspace}  
\newcommand{\pro}{P\xspace}  
\newcommand{\tidx}{r\xspace}  
\newcommand{\classw}{w\xspace}  
\newcommand{\classi}{k\xspace}  

\newcommand{\word}{v\xspace}  

\newcommand{\WordClass}{WordClass\xspace}

\newcommand{\Words}{W\xspace}  
\newcommand{\Dict}{D\xspace}  

\newcommand{\WS}{WS\xspace}  

\newcommand{\TofI}{malicious threads\xspace} 

\newcommand{\specificaiton}{specialization \xspace}
\newcommand{\generalization}{Generalization\xspace}

%
\begin{abstract}
How can we get a security forum to ``tell" us its activities and events of interest? 
We take a unique angle: we want to identify these activities without any a priori knowledge, which is a key difference compared to most of the previous problem formulations.
Despite some recent efforts, mining security forums to extract useful information has received relatively little attention, while most of them are usually searching for specific information.
We propose \algo,  an unsupervised tensor-based approach, to  systematically
identify important events in a three-dimensional space: (a) user, (b) thread, and (c) time.
Our method consists of three high-level steps:
(a) a tensor-based clustering across the three dimensions,
(b) an extensive cluster profiling  that uses both content and behavioral features,
and (c) a deeper investigation, where we identify key users and threads  within the events of interest.
In addition, we implement our approach as a powerful and easy-to-use platform
for practitioners. 
In our evaluation, we find that 83\% of our  clusters capture meaningful events
and 
we find more {\em meaningful} clusters compared to previous approaches.
Our approach and our platform constitute an important step towards detecting activities of interest from a forum in an unsupervised learning fashion in practice.

\end{abstract}


\pagestyle{plain}

\section{Introduction}


Security forums contain a wealth of information that  currently remains unexplored. Online security forums have  emerged as a platform where users generally initiate a discussion about their security-related issues. These forums aggregate valuable information in an unstructured way and initial work argues for a  wealth of useful information: emerging threats and attacks, promotion of hacking skills, and technical tutorials. Discussions around these topics at one or more points in time often involve a large number of users and threads, and we can think of them as important {\bf events} in the life of the forum.


How can we identify major events of interest in a forum in an unsupervised fashion?
The input is  a forum, and the desired outputs are the key events that capture the main activity in the forum and could be of interest to a security  analyst. For example, a security analyst would want to identify  outbreaks of attacks, the emergence of new technologies, groups of hackers with tight and focused interests, and underground black markets of hacking services. The challenges are that the information is unstructured and that we want to do this in an unsupervised way: {\bf we want the forum to \enquote{tell} us its events of interest}.

Mining security forums  have received relatively little attention and only recently.
We can identify three main categories of related efforts:
(a) security forum studies,
(b) analysis of blogs, social media, and other types of forums, and
(c) tensor-based mining approaches.
We discuss these efforts in Section \ref{sec:related}.


\begin{figure}[h]
    \centering
    \includegraphics[width=1\linewidth]{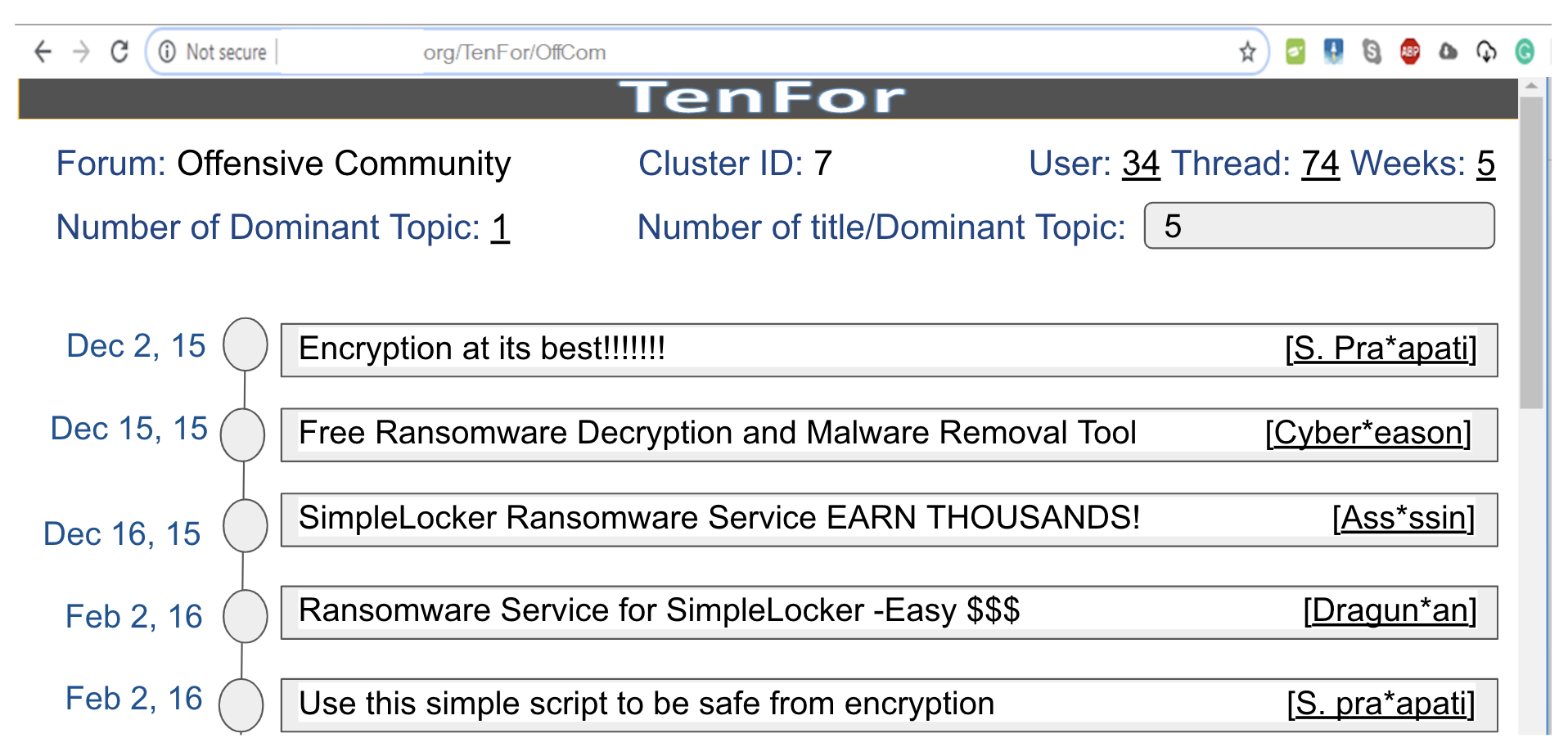}
    \caption{\storyline: a user-friendly visualization of a cluster summary with our \algo tool. We present one of the identified clusters, which captures the emergence of SimpleLocker ransomware from the Offensive Community forum. }
    \label{fig:event_report}
    \vspace{-4mm}
\end{figure}

As our key contribution, we propose, \algo,  a systematic tensor-based approach and tool to 
identify important events in an unsupervised way in a forum. Our approach operates at the three-dimensional space of (a) users, (b) threads, and (c) time.
Our method consists of three main steps:
(a) {\it clustering} using a tensor decomposition,
(b) {\it profiling} using both content and behavioral metrics, and 
(c) {\it investigating} using an automated,
but customizable, method to capture the dynamics of the cluster 
and provide an interpretable view.


We can summarize the  novelty of our approach in terms of techniques and features as follows:
(a) it operates in an unsupervised way, 
(b) it adapts and combines tensor-based clustering, behavioral profiling, and NLP methods,
(c) it is user-friendly by being parameter-free with {\it optional} user-specified ``knobs” that can adjust 
the granularity and information detail of the results, and
(d) it provides visual and intuitive fingerprints
of the events of interest.
All these capabilities are further discussed later.


     
     

Overall, an end-user can obtain the following results: (a) the most dominant clusters in the lifespan of the forum, (b) profiling information about these clusters including key users, key threads, key dates, and key topics and  keywords,
    (c) optional labeling of the clusters using user-specified  keywords.
Visually, the results can appear in a \storyline or a \tableview as shown in Fig.\ref{fig:event_report} and Table \ref{tab:1result} respectively.



In our evaluation, we apply \algo
on three security forums and one gaming forum with a total of 54000 users. 
We find that {\bf 83\%} of our identified clusters revolve around interesting events and each cluster shows 
high intra-cluster thread similarity,
as validated by experts, crowdsourcing, and other methods. Our approach also compares favorably with previous approaches~\cite{shah2015timecrunch, alzahrani2016community} leveraging the power of tensor decomposition to  strike the balance between size and number of clusters.

{\bf Going beyond security forums.} Although we focus primarily on security forums here, \algo can be used on other types of forums. As a proof of this, we apply our
approach on an online gaming forum and find interesting  activities,  including revenge hacking and romance scamming, as we discuss later.


{\bf The overarching vision.} 
 As a tangible contribution, we develop a powerful user-friendly platform that will be useful to both researchers, and industry practitioners.
Our ambition is to make this platform a reference tool for forum analysis and inspire subsequent research and development.\footnote{{\fontfamily{qcr}\selectfont Sample code: https://github.com/RisulIslam/TenFor}}
The proposed hands-free event extraction
is a significant  capability: we let the forum to tell us what are the key activities of interest, namely  ``taking the pulse" of the forum. This can enable practitioners to shift through a large number of forums of interest efficiently and effectively.
In the future, we will extend our tool by providing additional user-centric and content-centric capabilities.




\section{Datasets and Terminology}
\label{sec:data}

\begin{table}[t]
    \caption{Basic statistics of our forums.}
    \footnotesize
    \centering
    \begin{tabular}{|P{0.27\linewidth}|P{0.07\linewidth}|P{0.12\linewidth}|P{0.09\linewidth}|P{0.19\linewidth}|}
         \hline
         \textbf{Forum} & \textbf{Users} & \textbf{Threads} &  \textbf{Posts} & \textbf{Active Days}\\
         \hline
         Offensive Comm. & 5412 & 3214 & 23918 & 1239\\
         \hline
         Ethical Hacker &  5482 & 3290 & 22434 & 1175\\
         \hline
         Hack This Site & 2970 & 2740 & 20116 & 982 \\
         \hline
         MPGH & 37001 & 49343 & 100000 & 289 \\
         \hline
    \end{tabular}
    \label{tab:stat}
    \vspace{-4mm}
\end{table}

We apply our method on four forums in our archive, which consists of three security forums: Offensive Community (OC), Ethical Hacker (EH), Hack This Site (HTS), and one online gaming forum: Multi-Player Game Hacking Cheats (MPGH)~\cite{secforums}. All of these forums are in English language.

The dataset of the security forums spans 5 years from 2013 to 2017. Users in these forums initiate  security-related discussion  threads  in  which  other interested users  can  post  to  share  their  opinion. The threads fall in the grey area, mainly discussing both ``black-hat" and ``white-hat" skills.

MPGH is one of the largest online gaming communities with millions of discussions regarding different insider tricks, cheats, strategy, and group formation for different online games. The dataset was collected for 2018 and contains 100K comments of 37K users~\cite{pastrana2018crimebb}.

For completeness, we start with some terminology. Each {\em thread} has a {\it title} and is started by its first post, and we refer to subsequent posts as {\it comments}. 
The \textit{duration} of a thread is defined by the time difference between the first and last post of that thread. The {\em active days} for a forum are the number of days when the dataset contains at least one post. Each tuple in our dataset maintains the following format, $F$:=(forum ID, thread ID, post ID, username, date, and post content). The basic statistics of the data are shown in Table \ref{tab:stat}.
\section{Our Approach}
\label{sec:approach}

\begin{figure}[t]
    \centering
    \includegraphics[width=7.2cm, height=4.25cm]{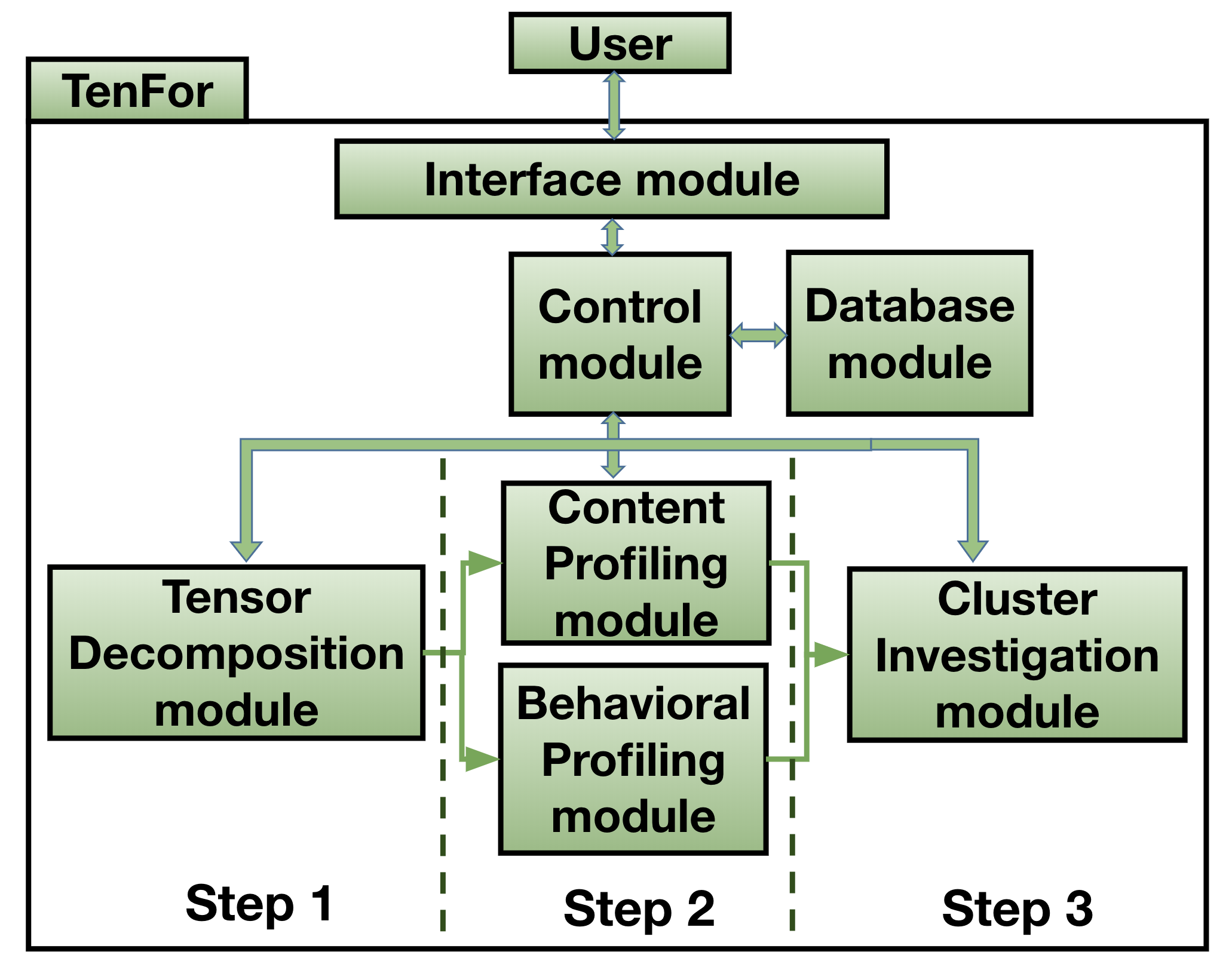}
    \caption{Overview of the  \algo approach and its  steps:  Step 1: {\bf Cluster};  Step 2: {\bf Profile};  Step 3: {\bf Investigate}.}
    \label{fig:ourapproach}
    \vspace{-4mm}
\end{figure}

We present, \algo,  a tensor-based multi-step approach, that identifies events and activities  in an unsupervised way.  
Fig. \ref{fig:ourapproach} provides
  the architecture of the platform. The Control module  communicates with Interface and Database modules. 
  The algorithmic core is provided by the Tensor Decomposition, Content Profiling, Behavioral Profiling, and Investigation modules.


We present an overview of  \algo, which works in three steps: a)  clustering via tensor-based decomposition, b)  cluster profiling, and c) cluster investigation as shown in Fig.~\ref{fig:ourapproach}.

{\bf Automated operation with optional user control.} A key design principle of our approach is to operate parameter-free, and at the same time, provide the end-users with ``knobs" for obtaining results of interest. 
Naturally, a savvy end-user can exert even more control by specifying algorithmic parameters
with well-defined APIs, especially in the tensor decomposition, which we discuss below. We revisit these parameters at the end of this section.


\subsection{Step 1: Tensor-based clustering} 


We provide an overview of the challenges  and algorithmic choices in our approach, starting with an introduction to tensors and tensor decomposition. 

\begin{figure}[t]
    \centering
    \includegraphics[ width=0.85\linewidth]{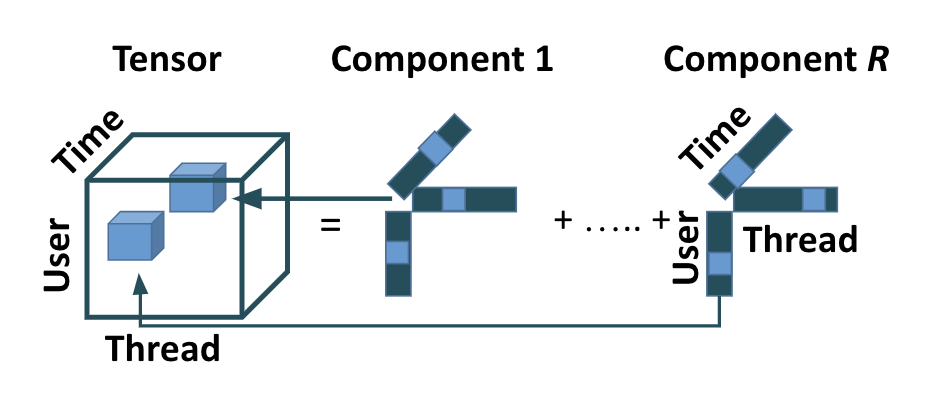}
    \caption{Visualization of tensor decomposition.}
    \label{fig:2tensor}
    \vspace{-6mm}
\end{figure}

{\bf Tensors and decomposition.} A d-mode tensor~\cite{kolda2009tensor} is a d-way array (here we use $d=3$). So, we call \scalebox{0.85}{$I \times J \times K$} tensor a  ``3-mode'' tensor where ``modes'' are the fixed number of indices to index the tensor; for us the ``modes'' being the user (U), thread (T), and weekly discretized time (W). Each 3D element of the tensor, \textit{X(i,j,k)}, captures the total number of interaction (in terms of \#comments) of user \textit{i} in thread \textit{j} at  discretized week \textit{k} or zero in the absence of such interaction. 
In a decomposition, we  decompose a tensor into \textit{R} rank-one components, where \textit{R} is the rank of the tensor, as shown in Fig.~\ref{fig:2tensor}. That means tensor is factorized into a sum of rank-one tensors i.e. a sum of outer products of three vectors (for three modes):  \scalebox{0.85}{$X \approx \sum_{r=1}^{r=R} U(:,r)\circ T(:,r)	\circ W(:,r)$} where \scalebox{0.85}{$U \in \mathbf{R}^{I \times R}$ , $T \in \mathbf{R}^{J \times R}$, $W \in \mathbf{R}^{k \times R}$} and the outer product is derived by \scalebox{0.85}{$(U(:,r)	\circ T(:,r)$ $\circ W(:,r))(i,j,k)$ = $U(i,r)T(j,r)W(k,r)$} for all \textit{i, j, k}.
Each component represents a latent pattern in the data, and we refer to it as {\bf cluster}.
For example, one such cluster in OC  represents a group of 29 users that are active in the first weekends of July 2016 and discuss ``multi-factor authentication failure" in a group of 72 threads.
Each cluster is defined by three vectors, one for each dimension, which show the \enquote{participation strength} of each element for that cluster.
Typically, one considers a threshold to filter out elements that do not exhibit significant \enquote{participation strength}, as we discuss later.

We need to address the following challenges to make the tensor decomposition work well in our domain.

\textit{a. What is the ideal number of components to target in the decomposition?} To answer this question, we use the AutoTen method~\cite{papalexakis2016automatic} and find the rank (R) of the tensor, which points to the ideal number of clusters to be decomposed into. AutoTen uses the {\it Core Consistency Diagnostic} metric in {\fontfamily{qcr}\selectfont CP\_ALS} and {\fontfamily{qcr}\selectfont CP\_APR} to find two probable ranks and finally chooses the max rank for the decomposition. So, the final rank, $R$, of a tensor is computed as follows: $R = max(R_{{\text{\fontfamily{qcr}\selectfont CP\_ALS}}}, R_{{\text{\fontfamily{qcr}\selectfont CP\_APR}}})$.

\textit{b. How can we decompose the tensor?}  We use the Canonical Polyadic or CANDECOMP/ PARAFAC (CP) decomposition to find the clusters.
The factorization may contain negative numbers in the decomposed components whereas our strategy of capturing the interaction between users and threads at different times is inherently non-negative.  We can achieve the non-negative factorization by adding the non-negative constraint in CP decomposition.

\begin{figure}[t]
    \centering
    \includegraphics[height=4cm, width=0.85\linewidth]{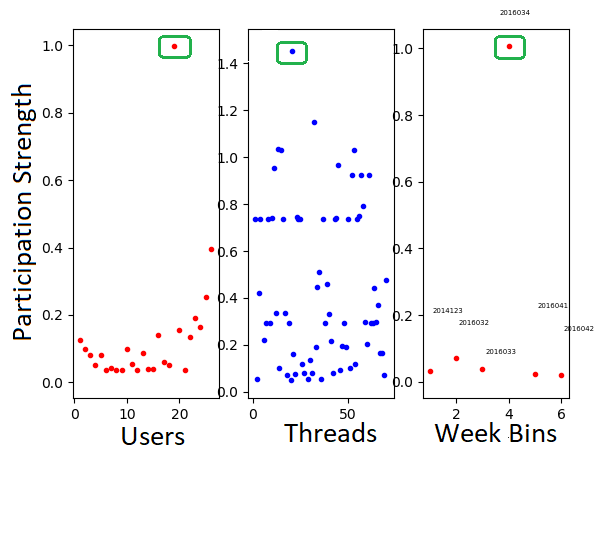}
        \caption{An example of a cluster (28 Users, 70 Threads, 6 Weeks) from OC. The intensity in each vector helps us identify users, threads and time intervals that are ``important" for the cluster. }
    \label{fig:4cluster_example}
    \vspace{-6mm}
\end{figure}

\textit{c. How can we strike a balance on cluster size?}
Each cluster is defined by three vectors (user, thread, and time), whose lengths are equal to the dimensions of the tensor as shown in Fig.~\ref{fig:2tensor}. We need a threshold to determine \enquote{significant participation in the cluster}, which is a common practice for (a) avoiding unreasonably dense clusters~\cite{sapienza2018non}, (b) enhancing interpretability, and (c) suppressing noise.
So, the challenge is to impose this sparsity constraint and eliminate the need for ad-hoc thresholding to find the clusters with only significant users, threads, and times. 
Our solution is to add $L_1$ norm regularization with non-negative CP decomposition. $L_1$ regularization pushes the small non-zero values towards zero.
Therefore, for each vector, we filter out the zero-valued elements and produce clusters with significant users, threads, and weeks only. In this way, we can eliminate the noisy users, threads, and weeks having the least significant contributions in the forum. The final model that we use for finding the clusters looks like this:

\scalebox{0.70}{%
$ \underset{U\geq 0, T\geq 0, W\geq 0}{\text{min}} \left\lVert X-D \right\rVert_{F}^{2} + \lambda ( \sum_{i,r} |U(i,r)| + \sum_{j,r} |T(j,r)| + \sum_{k,r} |W(k,r)|) $
}
where $\lambda$ is the sparsity regularizer penalty (set to 1) and \scalebox{0.75}{$D = \sum_{r} U(:,r)\circ T(:,r)\circ W(:,r)$}. To find the clusters, we solve the above equation. Since the equation is highly non-convex in nature, we use the well-established Alternating Least Squares (ALS) optimizer as the solver.
An example of a cluster after filtering is shown in Fig.~\ref{fig:4cluster_example} and is further discussed  in Section \ref{sec:results}.


\subsection{Step 2: Profiling the clusters} 
Having obtained the clusters, we propose to use content-based and behavior-based profiling to provide information and context for each cluster.

\textbf{Step 2.1. Content-Based Profiling:} We propose to profile clusters using content with the aid of two interconnected steps.

{\bf a. Cluster characterization:}
We identify the top {\it N}  keywords using TF-IDF from the first post of each thread in each cluster. Prior work argues that  the first post of a thread captures the focus of the thread~\cite{Joobin}. We use the term {\em cluster keywords} to refer to this set of words. These keywords can already provide a feel for the nature of the cluster,
but we also use more sophisticated techniques in the next step.

\textbf{b. Cluster labeling:} We give the end-user the ability to define classes of interest that we then use to label the clusters. 
For ease of use, the end-users can define a class by providing 
 a bag of words. To label the clusters,
 we compare these bags of words with the {\em cluster keywords} from the previous step.

To demonstrate this capability, 
 we start with a group of classes that would be of interest to a security analyst.
 Specifically, we adopt the following classes of interest from  prior work~\cite{Joobin},
 which defines four types of threads:
{\bf Announcement type (A)}  where people announce news and events, including  hacking achievements and cyber-attacks; {\bf Product type (P)},  where people buy or sell hacking services and tools; {\bf Tutorial type (T),}  where users post tutorials on how to secure or hack into systems;  and,  {\bf General Discussion type (G)}, which is the category for all threads not in the above categories. 

We then  calculate how ``relevant" each cluster is to each class type. 
For consistency, we have adopted the 
same definitions
for these categories as the aforementioned work.
To do this,
we compute the Jaccard Similarity
between the \textit{cluster keywords} and the keywords that define each class type. We  label the cluster as \textit{A, T, P, G} type based on the highest Jaccard Similarity score. A cluster can be labeled as \textbf{Mix} type if the similarity scores of different types are within a close range (defined as $\pm 0.02$). 


{\bf Step 2.2. Behavior profiling:} 
To provide more information per cluster,
we  use behavioral properties, which capture how users and threads interact with each other over time.
We provide the following groups of capabilities and plots to the end-users:

\textbf{a.} \textbf{Basic Distribution plots} of metrics of the clusters in a forum, such as the distribution of \#users, \#threads, \#active days etc. per cluster of the forum.

\textbf{b.} \textbf{Scree-plots} of metrics of clusters, which capture the pair-wise relationships of different metrics of clusters, 
such as \#threads vs \#users, \% of active days vs duration (defined as the time difference between the last and the first post of the cluster) for each cluster of the forum as shown in Fig.~\ref{fig:screeuserthread} and \ref{fig:screedurationactivedays}. 

\textbf{c.} \textbf{Heat map visualizations} of the clusters and the relative strength of their behavioral metrics.
Currently, we use ten behavioral metrics that include
the average (over the cluster): average post length per user, number of threads initiated per user, comment to thread participation ratio of the users,
number of comments per user, number of active days of the threads etc.
We normalize the values of the averaged  metrics and present the behavioral profiles  using a heat-map-style plot 
 as we show, and discuss later, in Fig.~\ref{fig:5profile_cluster}.
 
The visual depiction helps an analyst to quickly gauge the behavioral profile of the clusters and spot differences. Also, we expand this functionality by developing an automated capability to report the anomalous cluster/s using standard DBScan anomaly detection algorithm \cite{sheridan2020application} in these profiles. We discuss the findings in Section \ref{sec:results}. 

\subsection{Step 3: Investigation of clusters}

We develop a suite of capabilities that can help automate
an  in-depth investigation of the clusters coming from the previous steps.
Although this can be done manually, the goal is to
make the life of an analyst easier. 
Our platform provides the user with well-organized and easily accessible information trying to strike the balance between
being informative and intuitively interpretable.
Moreover, we develop two ways so that the end-users can summarize the clusters: (i) \storyline, and (ii) \tableview.

{\bf Step 3.1. Creating the \storyline:} We develop a systematic and, arguably, more interpretable method to 
capture the essence of a cluster by highlighting the {\it k} most indicative  threads in 
a non-decreasing temporal order as shown  in Fig.~\ref{fig:event_report}. To accomplish this, we follow the process described below.


Identifying the important threads for the cluster is calculated in the following stages.
In stage one, we find an extended list of topics, $T_{ext}$, for the whole cluster. To do this, we use the commonly-used LDA Bag-of-Words model \cite{sharma2017survey}, and  we focus on the {\em titles} of the threads in the cluster threads because the {\em titles} provide a compact and meaningful summary of the threads. 
In stage two,  we calculate the {\it relevance scores} of each thread with respect to each topic $t \in T_{ext}$. 
We associate each thread with the topic with the highest {\it relevance score}.
In stage three, we find the most representing topics, $T_{dom}$, of the cluster. To achieve this, we find the distribution of the number of threads per topic in the decreasing order and from there we choose the  list of \textit{dominant topics, $T_{dom}$},
which we define as the minimum number 
of topics that represent at least ``thread threshold", $Th_{dom}$=70\%(default) of the threads.
In stage four,  we identify the top \RelThreads most relevant threads  based on their relevance score for each of the dominant topics in $T_{dom}$. We then present them in a non-decreasing temporal order as shown in Fig.~\ref{fig:event_report}. Note that the parameter 
\RelThreads has a default value of 5, but the end-user can adjust it to her liking.

Here, we focus on the {\em titles} as we want to have the title of thread ``tell the story" in a visceral and intuitive way for the end-users. In the future, we will consider the text of the whole thread to find topics and consider more involved topic extraction methods.

{\bf Step 3.2. Creating the \tableview:}
We provide an alternative way to view all the clusters 
in the forum in a way that puts emphasis on key authors and key threads. 
 This \tableview can provide compact event summarization and key entities in each cluster.
 We argue that this may be appealing
for a different type of analysis.
 Table~\ref{tab:1result} demonstrates the \tableview that we provide.
In our platform, we have clickable links that one can follow to investigate these entities of interest providing an interactive capability. We now present the generation of the columns of Table~\ref{tab:1result}.

{\bf a. Identifying important entities: users, threads, and time intervals.}
We propose a method to identify the most dominant users, threads, and time periods, where significant activity takes place and populate the columns 4, 5, and 6 in Table~\ref{tab:1result}. Specifically, we propose to identify the top \textit{k} entities from each cluster, where $k \geq 1$ with the default being \textit{k}=3. 
We use the factorized vectors to gauge the ``importance" of an entity in a cluster as shown in Fig.~\ref{fig:4cluster_example}. 
The  green boxes show the entities with the highest ``Participation Strength''. From each of the top {\it k} weeks, we also report the most active day in terms of the highest \#post made in that week.

Note that the parameter \textit{k} can be modified by the end-user
to adjust to her preference or type of investigation.


{\bf b. Representing the nature of the cluster in \tableview:}
We present another way to capture the essence of a cluster,
which we provide as text in the last column of Table~\ref{tab:1result}. 
Obviously, there are many ways to achieve this.
We opt to report the most dominant topics, $T_{dom}$, per cluster which is a common practice to represent and interpret events~\cite{shi2019human,mehta2019event}.
We have already discussed a method to identify the dominant topics above, which can be provided in the final column in Table~\ref{tab:1result}. 
Note that in Table~\ref{tab:1result}, we start from the dominant topics,
but reconstruct the events within each cluster to provide more context to the readers.


{\bf The optionally tunable parameters of \algo:} \algo can operate without any user input, but we expose the following parameters to a savvy user who wants to experiment, which we list here along with their default values: (a) temporal granularity: week, (b) size of the cluster keywords {\it N: 50},
(c) cluster labels: A, T, P, G/Mix as defined here, (d) thread threshold for dominant topic $Th_{dom}$=70\%, (e) \#relevant threads in \storyline $R_t$: 5, and (f) \#top entities in \tableview k: 3.
\section{Results and Evaluation}
\label{sec:results}

\begin{figure*}[t]
    \begin{subfigure}{0.28\textwidth}
    \includegraphics[height=3cm,width=6cm]{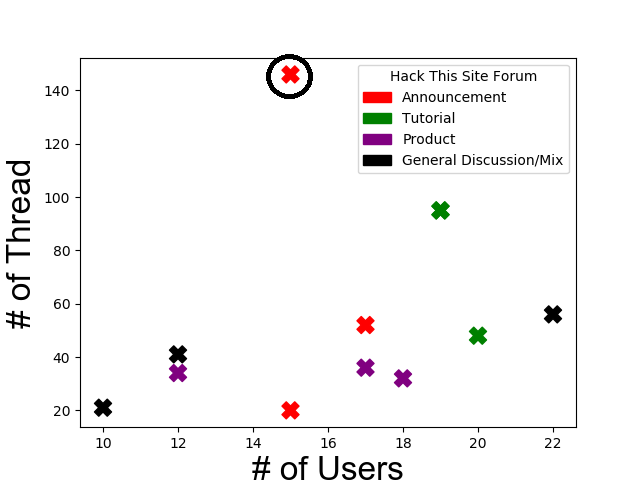}
        \caption{Scree plots of \#users vs \#threads in A, T, P, Mix/G type clusters of HTS.}
    \label{fig:screeuserthread}
    \end{subfigure}
    ~
    \begin{subfigure}{0.30\textwidth}
    \includegraphics[height=3cm,width=6cm]{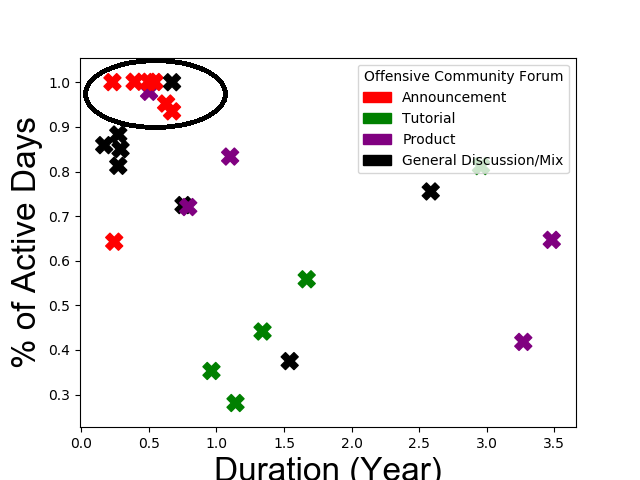}
        \caption{Scree plots of \% of Active Days vs Duration in A, T, P, Mix/G type clusters of OC.}
    \label{fig:screedurationactivedays}
    \end{subfigure}
    ~
    \begin{subfigure}{0.37\textwidth}
    \includegraphics[height=3cm,width=7cm]{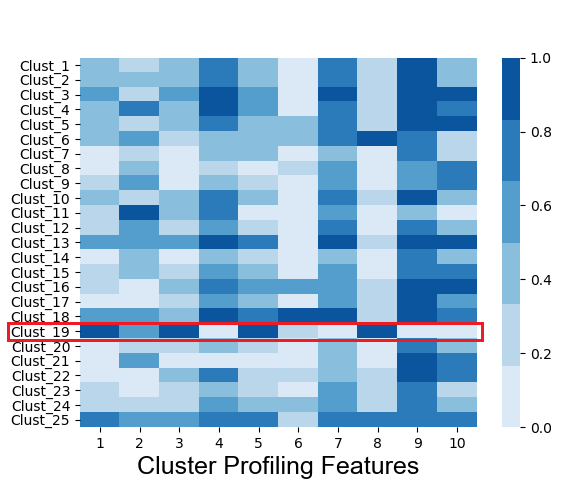}
    \caption{Behavioral profiling of the clusters from OC: x-axis is  different behavioral features, and y-axis is cluster IDs. Case-study: Cluster 19 has a unique feature intensity profile.}
    \label{fig:5profile_cluster}
    \end{subfigure}
    
    \caption{Behavioral profiling of the clusters. 
    }
    \label{fig:profiling}
    \vspace{-2mm}
\end{figure*}

\begin{table*}[t]
    
    \caption{Properties of the clusters in OC, HTS and EH. Here U=user, Th=thread, W=weeks and the percentage of users, threads in a particular type of cluster is based on total number of users, threads in each forum. 
    } 
    \footnotesize
    \centering
   \begin{tabular}{|c|c|c|c|c|c|c|c|c|c|c|c|c|c|c|c|c|c|}
        \hline
        \centering
        \multirow{2}{*}{\textbf{Forum}} & \multicolumn{3}{p{0.12\linewidth}|}{\textbf{Initial Tensor Size}} & \multicolumn{3}{p{0.19\linewidth}|}{\textbf{filtered Entities in clusters}} &
        \multicolumn{1}{p{0.05\linewidth}|}{\textbf{\#Cluster }} & 
        \multicolumn{4}{p{0.14\linewidth}|}{\textbf{\#Cluster  per Type }} &  
        \multicolumn{2}{p{0.08\linewidth}|}{\textbf{Type A (\%) }}& 
        \multicolumn{2}{p{0.08\linewidth}|}{\textbf{Type \textbf{T} (\%) }}&  
        \multicolumn{2}{p{0.08\linewidth}|}{\textbf{Type \textbf{P}   (\%)}}\\
        \cline{2-7}
        \cline{9-18}
        & U& Th& W& U& Th& W& &  A& T& P & Mix/G &U & Th& U& Th& U& Th\\
        \hline
        \centering
        OC& 5412& 3214& 240& 1086& 2505& 107& 25& 7& 5& 5& 8& 5.7& 21.2 & 5.1& 14.3& 3.8 & 14.3 \\
        \hline
        \centering
        HTS& 2970& 2740& 240& 196& 676& 59 & 12& 3& 3& 3 & 3 &1.5&7.9 & 2& 3.7 & 1.6& 3.7\\
        \hline
        \centering
        EH& 5482& 3290& 240 & 315& 424& 82 & 15 &3 & 6 & 2 & 4 & 1.1 &2.3  & 2.8& 2.2 & 0.6 & 1.3\\
        \hline
    \end{tabular}
    
    \label{tab:clustsummary}
    \vspace{-4mm}
\end{table*}

We apply our method on four forums in our archive discussed in Section \ref{sec:data}. We discuss the output of each step of \algo for three security forums below. 

\subsection{Step by step output provided by \algo}
 
{\bf Step 1.} We do a tensor decomposition for each forum. 
We provide an overview of the results of our decomposition in  Table~\ref{tab:clustsummary}.
Note that we opt to use \textit{week} as the unit of time,
but we experimented with  days and months.
Since we want to capture events, a week seems to strike a good balance between a day, and a month, which could be too short and too long respectively. 
We  find  the target number of clusters with the method described earlier. We get a total of 52 clusters from all three forums of which 25 clusters from OC, 12 clusters from HTS, and 15 clusters from EH. Note that we did experiment with more clusters than the ideal number, but that  yielded extremely small clusters (e.g. 2 users, 3 threads, 1 week).

{\bf Step 2a. Content-based profiling and labeling.} We use the A, T, P, or Mix/G labels, which we defined earlier. We set the \#  {\it cluster keywords, N}=50.
Note that we report Mix and G types together here for the ease of presentation.

An overview of the clusters and their properties for all three forums is presented in Table~\ref{tab:clustsummary}. 
Specifically, we find the following distribution of clusters: (a) {\bf 26\% of the clusters correspond to real security events}, such as attacks, (b) {\bf 22\% of them represent black market communities} for malware tools and services, and (c) {\bf 32\% of them represent security tutorials, events, and communities}, with most tutorials sharing malware and penetration techniques.

{\bf Step 2b. Behavior profiling.}
We provide the functionality to profile clusters
based on their activity and dynamics.
Apart from providing a general understanding,
the analysis can help us  spot outliers, which the end-users can investigate in Step 3. 

First, our  \algo platform provides some basic distribution plots, scree plots, and a heat map of the behavioral-based profile for each cluster, as described earlier. Due to space limitations, we only show two indicative scree plots in Fig. \ref{fig:screeuserthread} and \ref{fig:screedurationactivedays}.
In Fig. \ref{fig:screeuserthread} for HTS, the black-circled cluster at the top is an \textbf{A} cluster, where just 15 people participate in a comparatively huge number of 145 threads. Upon further inspection, they are a group of hackers boasting about their hacking success. Some indicative {\it cluster keywords} of this cluster are {\it hack, brag, success, breach }etc. (provided by \algo in wordcloud format as well) which actually substantiate our claim.

Similarly, in Fig.~\ref{fig:screedurationactivedays} for the OC forum, the encircled clusters exhibit continuous activity: the Percentage of Active Days over the Duration of the cluster is more than 90\%! 
This is an indication of an ``urgency" in the cluster when compared with the typically lower Percentage of Active Days. This urgency 
is amply illustrated by cluster 9 (22 users, 60 threads): 
users talk about ``strike week", during which the government attacked organized cyber-crime in the second half of March 2015. Strike week created frantic activity in the forum at that time.    

\begin{table*}[t]
    
    \caption{Investigating nine clusters identified by \algo reveals interesting activities. (CID is the id of the cluster).
    } 
    \footnotesize
    \centering
  \begin{tabular}{|P{0.05\linewidth}|P{0.032\linewidth}|P{0.029\linewidth}|p{0.09\linewidth}|p{0.1\linewidth}|p{0.10\linewidth}|p{0.43\linewidth}|}
        
        \hline
         \textbf{Forum-CID} & \textbf{No. Users} & \textbf{Type} & \textbf{Top Threads} & \textbf{Top  Users} & \textbf{Top  Dates} & \textbf{Events and Explanation}
         \\
        \hline
        OC-7 & 34 & P & 3502, 4843, 4841 & S. Prajapati 23, Cyberseason, Assassin & Dec  2 \& 15 2015,  Feb 13 2016 & (a) A market of 34 buyer/sellers of decryption tools against SimpleLocker ransomware with peaks in Dec 2015 and again in Feb 2016, which mirrors the outbreak events of SimpleLocker. \\
        \hline
        OC-8 & 54 & A & 2562, 1228, 1234 & V4nD4l, RF, Pratham & Feb 4, 19, \& 28 2016 & A peak is detected: a) when V4nD4L claimed success in hacking Facebook in Feb, 2016, b) V4nD4L recruits seven members in a hacking group.\\
        \hline
        OC-12 &42 & T & 804, 6995, 2099 & Dragunman, Pratham, L1nkm3n & June 4, 9 \& 19 2015 & (a) Five people collaborated to  spread the RAT virus; (b) Dragunman shared tutorials on hacking into banks; (c)  Pratham promoted several YouTube videos on hacking WiFi networks.\\
        \hline
        HTS-3 & 63 & A & 890, 10594, 11349 & TheMindRapist, cdrain, Bhaal & Oct 5 \& 12, Nov 22 2013 & TheMindRapist announced a hacking web-platform where people can submit the URLs they want to have hacked in Oct 5, 2013.\\
        \hline
        HTS-6& 39 & T & 1125, 234, 6788 & Ninjex, Rajor, mShred & Apr 7, Aug 22 \& 31 2014 & A peak in activities is observed when Ninjex and mShred shared tutorials for 
        building hacking tools during the reported Top Dates.\\
        \hline
        HTS-12 & 18 & P & 3453, 4467, 8901 & whacker, DoSman, Bhaal  & April 10 \& 28, May 12 2016 & DoSman offered a 30 days free trial of a  DoS attack tool with a peak in April, 2016.\\
        \hline
        EH-2 &31 & A & 7263, 8762, 9127 & DarkKnight, Don, VandaDGod  & Feb 1 \& 9 2016, May 15 2017 & DarKnight was a victim of Locky ransomware in Feb 2016, which sparked a large discussion. Also WannaCry ransomware created a huge fuss in security world in May 2017\\
        \hline
        EH-3 & 26 & P & 4563, 213, 4498 & hayabusa, dynamik, azmatt  & May 12 2017, July 3 2017, Dec 2 2017 & The peak at the Top Days is due to  hayabusa and azmatt offering to sell malware tools:  xchat tool for windows, hidden surveillance tools, webcam hacking tool etc.\\
        \hline
        EH-6 &46 & T & 1251, 8325, 8338 & Don, D3vil, VandaDGod  & Nov 19 \& 24 2017  & VandaDGod, a expert Linux hacker, shared a popular tutorial series of hacking in Kali Linux in Nov 2017.\\
        \hline
    \end{tabular}
    
    \label{tab:1result}
    \vspace{-4mm}
\end{table*}

Finally, we also provide a compact visual behavioral profile for each cluster shown in Fig.~\ref{fig:5profile_cluster} for the OC forum.  This can convey condensed information to the end-users visually.
For example,  cluster 19 (40 users, 88 threads), highlighted with the red box,  seems to have a rare combination of active (dark blue) features. Specifically, these features suggest that the cluster exhibits high values of (a) average length of the first post of a thread per user (feature 1),  (b) average ratio of \#comments to \#threads which a user generates or participates in (feature 3),  and  (c) average \#comments per thread (feature 5). This behavior of the cluster is aligned with a \textit{Tutorial} type cluster:
 (a) the first post is usually long, 
 (b) tutorials often spark discussions, leading to multiple comments by a user in a thread, and (c) there are many questions and ``thank you" comments in a tutorial thread.
Note that this is also the label that our content-based labeling suggests.


{\bf Step 3.} We showcase how  we can enable a deeper analysis for each cluster with (a) \tableview, and (b) \storyline. An example of our \tableview is presented in Table~\ref{tab:1result} where we highlight three selected clusters from each forum and we provide the information in terms of the type of the cluster, most significant threads, users, and dates.
The final column is populated with the {\it dominant topics}, though here, we provide a manually-enhanced 
reconstruction of events for presentation purposes.
As explained earlier, we also present a \storyline where we identify the top-{\it k} most indicative threads of the cluster which provides a human-readable thumbprint of the cluster. In Fig. \ref{fig:event_report}, we show such a result that was generated automatically for cluster 7 (34 users, 125 threads) of the OC forum. We find that one topic, ransomware, represents 81\% of the {\it titles}. In default settings, \algo reports top {\it k=5} {\it titles} based on the highest \textit{relevance score} for the ``ransomware'' topic in a sorted timeline fashion. From this \storyline, the analyst can easily come into a conclusion that the cluster actually captures the spread of SimpleLocker. Therefore, this view is particularly useful for clusters that capture an event or a discussion, as they can provide the evolution of the event as captured by its most dominant threads.


We discuss 9 of the clusters in Table~\ref{tab:1result} in more detail to show-case the kind of information that we can gain.

{\bf a. Detecting emerging security threats.} First,  several clusters consist of events that discuss novel security threats. For example, cluster 7 and 12 of OC revealed the growing concern of an 
extensive outbreak of the SimpleLocker ransomware and the RAT virus respectively. Also, cluster 2 of EH 
provides a timely warning of the  explosive outbreak of Locky ransomware in Feb 2016.

{\bf b. Identifying bad actors and their tools.} Our analysis can lead to important bad actors with Internet-wide reputation. 
Interestingly, it seems that hackers
use their usernames consistently around Internet forums, possibly enjoying their notoriety. For example, our analysis (also shown in Table~\ref{tab:1result}) leads to the usernames of hackers, ``V4ND4l", ``Dragunman" and ``VandaDGod". 
A simple Internet search of these usernames quickly leads to people with significant  hacking activities and hacking tutorials on YouTube offered by them.
Furthermore,  we find  that ``VandaDGod"  is active in multiple clusters in EH forum. 
In July 2019, a hacker group ``VandaTheGod" is reportedly accused of defacing dozens of government sites~\cite{vandaDgod}.

\subsection{Evaluation of \algo}
\label{subsec:man_evaluation}

Evaluating the effectiveness of our approach and tool
is inherently difficult due to the open-ended and  subjective nature of the problem.
We list our efforts to assess the precision and recall
of our approach by examining the
 precision and  recall to the best of our capabilities.
 

{\bf A. Precision.} We  present the evidence that our clusters are meaningful using several different angles.
We find that {\bf 83\%} of our clusters revolve around interesting events and each cluster shows high intra-cluster thread similarity. This is validated by a group of security experts and further corroborated via crowdsourcing and the REST methodology~\cite{Joobin}.

{\bf 1. Manual evaluation from domain experts. } 
We use a group of 3 security researchers to manually investigate all 52 clusters from all three forums. We asked the experts to (a) assign a  score (out of 100) for each cluster based on the topic cohesiveness, and (b) summarize the important event(s) in each cluster, if they think the topic cohesiveness score crosses 70. Our experts determined 43 clusters containing 55 significant events based on the majority vote.

{\bf 2. Manual evaluation via crowd-sourcing. } We recruited nine judges among graduate students across campus to check the 52 clusters whether they contain noteworthy events and assign a similar score per cluster  like the domain experts. A key difference is that our volunteers make their decisions based on 10 randomly selected  {\it thread titles} from each cluster. 
For calibrating their sensitivity, the judges were given two  sample clusters before the evaluation with (a) randomly selected thread titles, and (b) titles from the same topic.
Note that we declare a cluster as cohesive if at least five of the judges assign a topic cohesiveness score  $\geq$70.  The group declared 41/52 clusters (79\%)
as cohesive containing 56 events. For 52 clusters and 9 judges, we calculate the Fleiss' kappa score \cite{fleiss1973equivalence}, $\kappa=0.699$, which is substantial enough to come to a significant inter-annotator agreement in our context. 
In Table~\ref{tab:man_eval_percentage}, we provide an overview of the results above. 
We argue that each combination in Table~\ref{tab:man_eval_percentage} columns has its own merit with the intersection being the most strict and the union being the more inclusive.

{\bf 3. Assessing the cohesiveness.} 
We corroborate the effectiveness of our content-based labeling (as A, T, P, G type) and assess the cohesiveness of clusters in an indirect way using a state-of-the-art technique, REST~\cite{Joobin}.
REST follows a thread-centric  approach and labels threads along these four categories focusing on the content of a thread.
We applied REST for every thread in our clusters.
We find that 42 clusters have more than 70\% threads of the same type according to REST
and they also agree with our cluster label.
Note that REST operates at the level of a thread, while we label clusters, 
which will inevitably introduce ``errors". 
Thus, we consider the above matching numbers as a good indication for both: (a) the cohesiveness of our clusters,
and (b) the accuracy of our labeling approach.

Going one step further, we manually investigated the threads that REST was not confident enough to label.
We randomly selected 200 such threads and found that 81\% of these threads were aligned with the type of the clusters they were in.
Many of these threads were short, and we suspect that REST did not have enough context to assign a label.

{\bf B. Recall. } 
Quantifying the recall of our approach is even harder. As answering to
\textit{``Are we missing important activities and events?''} question is harder to prove, we attempt to argue in favor of our method by providing three types of observations.

{\bf 1. Any spike in activity is caught by \algo.} We argue that any event that creates significant activity involving threads and users will be caught by \algo. To provide evidence,  we find the top 20 weeks of high activity (in \# posts), and the top 50 active users and threads (in \# posts) in forum OC.
We find that 19 out of the 20 most active weeks, 47
out of the 50 most active users, and 46 out of the 50 most active threads are also identified among the top 5 ``performers" in our clusters (\textit{k=5}).

{\bf 2. Several real-life events are caught by \algo.}
\algo manages to capture several significant data breach events in the clusters from HTS forum including a) Sony Pictures, (b) Snapchat, and (c) Slack data breach. 
Users of security forums tend to be more interested in malware and ransomware  discussions. For example, among the six most widespread ransomware from 2013-2017 listed in \cite{Ransomware_List}, \algo captured 5 of them in 4 clusters: (a) SimpleLocker(2015-16) event in OC, (b) Locky(2016) and WannaCry(2017) ransomware event in EH, and (c) CryptoLocker(2014) and Petya(2016) ransomware in OC. Therefore, we argue that significant real-life events which  discussed  in the forums extensively are captured in the clusters.

{\bf C. Comparison with state-of-the-art methods.} We compare \algo with TimeCrunch \cite{shah2015timecrunch}, which identifies  temporal patterns in a dynamic graph. This is the closest state-of-the-art method: our input tensor can be seen as a dynamic bipartite graph. We argue that \algo is able to find more and meaningful cluster patterns compared to TimeCrunch. 

Specifically, applying the default parameter-free setting of TimeCrunch, we find a total of 17 temporal patterns from three security forums, whereas \algo finds a total of 52 clusters patterns. First,  upon further investigation, we find that 13 of these 17 temporal patterns are actually present in our identified clusters. TimeCrunch reports only  fixed types of patterns (full/near bipartite core, full/near clique, ranged/constant star etc.) based on Minimum Description Length (MDL) after encoding the model and the output patterns. Encoding larger clusters  leads to higher MDL cost, which may be why TimeCrunch reports clusters of smaller sizes. \algo does not consider any fixed types of pattern types and leverages the power of tensor decomposition.
Furthermore, we observe that all 17 clusters are small in size (less than 21 users), compared to the \algo cluster sizes (as much as 228 users).
It seems that TimeCrunch does not identify larger 
clusters- probably can not ``summarize" efficiently 
and, therefore, does not identify the interesting larger clusters which we show in Table III.



We also compare \algo with a  widely-used community finding algorithm for Weighted Bipartite Network (CFWBN)~\cite{alzahrani2016community}.
This approach operates on the user-thread space 
and identifies a total of 771 bipartite communities from all three forums.
However, we find 91\% of these clusters are small, with $\leq3$ users,
 and only 35 communities start becoming substantial with $\geq5$ users.
We argue that this large number of communities 
and the absence of time dimension make a follow-up investigation harder for the end-users.

In conclusion, \algo strikes a balance between reporting too many and too few meaningful clusters compared to previous other methods. Additionally, it provides the end-users with key actors and a timeline of key events in an informative visualization. 

\begin{table}[t]
    \caption{Precision of \algo: Percentage of clusters declared as interesting and cohesive in our evaluation.}
    \footnotesize
    \centering
    \begin{tabular}{|p{0.12\linewidth}|p{0.11\linewidth}|P{0.20\linewidth}|P{0.20\linewidth}||P{0.11\linewidth}|}
         \hline
         \textbf{Experts} & \textbf{Crowds} & \textbf{Expert AND Crowd} & \textbf{Expert OR Crowd}& \textbf{REST} \\
         \hline
         83\% & 79\% & 71\% &  88\%& 79\% \\
         \hline

    \end{tabular}
    \label{tab:man_eval_percentage}
    \vspace{-4mm}
\end{table}

{\bf D. Generalizability.} We wanted to see if our approach would work equally well on  different types of online forums of larger size.
For this reason,  we apply our method on our online gaming forum, MPGH, discussed in Section \ref{sec:data}. 
Applying \algo on this forum,
we find 41 clusters with a total of 1.3K users and 3K threads. Apart from finding clusters related to gaming strategy, and tricks for different popular online games, we also find several cyber-crime related activities even in this gaming forum! We highlight the indicative findings below.

{\em (a) Scamming and cheating.}
Interestingly, the biggest cluster with 300 users and 400 threads is focused solely on scamming. 
The key perpetrators are reported to be Nigerian scammers and a well-known scamming company, ``iYogi".

{\em (b) Romance scamming.}
We identify a sudden emergence of ``romance scamming" reports in the mid of August 2018. Apparently, scammers engage in online games, connect with other players, and win their affection and trust, which they use for monetary gain~\cite{romantic_scamming}. 

{\em (c) Hacking for hire.}
Another surprising behavior  is the search for a hacker
to exact revenge on a gaming rival, as captured in a cluster with 69 users and 119 threads.

Our initial results hint at a wealth of interesting behaviors in the gaming forums, which we will investigate in the future. 

{\bf E. Computational effort.} The computation required by \algo is not excessive. The average runtime for preparing the final \storyline of the biggest forum with 100K posts, MPGH, takes only 4.35 minutes on average. 
Our experiments were conducted on a
machine with 2.3GHz Intel Core i5 processor and 16GB RAM.
We use Python v3.6.3 packages to implement all the modules of \algo. 
We believe that the runtime can be reduced to seconds if we use more powerful hardware.
These results suggest that \algo scales reasonably well in practice. The sample code can be found at {\fontfamily{qcr}\selectfont https://github.com/RisulIslam/TenFor}.
\section{Related Work}
\label{sec:related}


Overall, none of the previous efforts combines: (a) using tensor decomposition, and (b) extracting events of interest in an unsupervised manner. The most related work to the best of our knowledge is TimeCrunch \cite{shah2015timecrunch}. TimeCrunch leverages the MDL principle and is limited to reporting only six fixed types of temporal patterns. It also does not use tensor decomposition and does not include a systematic event extraction mechanism like we do here.
We discuss other related works briefly below.

\textbf{a. Mining security forums:} 
Some recent studies focus on identifying key actors and emerging concerns  in security forums using supervised techniques and NLP  by utilizing their social and linguistics behavior~\cite{marin2018mining} 
Some of these works are empirical studies without developing  a systematic methodology.
Recent efforts include analyzing the dynamics of the black-market of hacking services~\cite{portnoff2017tools}, extracting  malicious IP addresses reported by users in security forums \cite{gharibshah2018ripex}. A recent work \cite{Joobin}, REST, identifies and classifies threads given keywords of interest, and we use it to validate our cluster labeling.

\textbf{b. Mining social networks and other types of forums:} Researchers have studied a wide range of online media such as blogs, commenting platforms, Reddit, Facebook etc. Some recent works  analyze the user behavioral patterns observed in Reddit~\cite{thukral2018analyzing} and infer information for the users from their  activities on Facebook \cite{bayer2018facebook} and GitHub~\cite{rokonSourcefinder2020}.
Despite some common algorithmic foundations, we argue that 
different media  and different questions require novel and targeted methods. 

Event detection is a broad and related type of research. Some recent works~\cite{shi2019human} focus on 
reporting the topics discussed by these influential users,  
real-time event detection from Twitter~\cite{hasan2019real}.
A recent work~\cite{mehta2019event} proposes a
hierarchical multi-aspect attention approach for event detection but does not consider the author and temporal dimension as we do here.


\textbf{c. Tensor Decomposition approaches:}
Tensors is a well-studied area with a wide range of diverse applications and domains ~\cite{kolda2009tensor} including understanding the multilingual social networks in online
immigrant communities~\cite{papalexakis2015understanding}, community assignment of nodes in multi-aspect graph \cite{gujral2018smacd}, 
and tensor-based community evolution~\cite{liu2019community}.
We are not aware of any tensor-based event extraction studies for online forums. In our work, we adapted the CP tensor decomposition \cite{kolda2009tensor, faber2003recent}  and combined it with L1 regularization to filter out the insignificant entities.


\section{Conclusion}

We propose, \algo,  an unsupervised-learning tensor-based approach to  systematically identify important events in a three-dimensional space: (a) users, (b) threads, and (c) time.
Our approach has three main advantages:
(a) it operates in an unsupervised way, 
though the user has ways to influence its focus, if so desired,
(b) it provides visual and intuitive information,
and 
(c) it  identifies both the events of interest, and the entities of interest within the event, including threads, users, and time intervals.

Our work is a step towards an automated unsupervised capability,
which can allow security analysts and researchers to shift through the wealth of information that exists in security forum and online forums in general.


\section{Acknowledgement}
This work was supported by the UC Multicampus-National Lab Collaborative Research and Training (UCNLCRT) award \#LFR18548554.

%

%
{ \bibliographystyle{IEEEtran}}
\bibliography{risul}

%

\end{document}